\documentclass[aps,prl,twocolumn]{revtex4}
\usepackage[dvips]{graphicx}

\begin{document}
\title{
Simulation study of the inhomogeneous Olami-Feder-Christensen model of earthquakes
}

\author{
 Takumi Yamamoto, Hajime Yoshino and Hikaru Kawamura
}

\affiliation{Department of Earth and Space Science, Faculty of Science,
Osaka University, Toyonaka 560-0043,
Japan}
\date{\today}
\begin{abstract}
Statistical properties of the inhomogeneous version of the Olami-Feder-Christensen (OFC) model of earthquakes is investigated by numerical simulations. The spatial inhomogeneity is assumed to be dynamical. Critical features found in the original homogeneous OFC model, {\it e.g.\/},  the Gutenberg-Richter law and the Omori law are often weakened or suppressed in the presence of inhomogeneity, whereas the characteristic features found in the original homogeneous OFC model, {\it e.g.\/},  the near-periodic recurrence of large events and the asperity-like phenomena persist.
\end{abstract}
\maketitle

 Statistical properties of earthquake have attracted much interest in seismology as well as in statistical physics \cite{Scholzbook,Kolkotta}. It has been known for years that power-laws are often observed there, {\it e.g.\/}, the Gutenberg-Richter(GR) law  associated with the earthquake size distribution or the Omori law associated with the time evolution of aftershock frequency, which leads to the view that earthquakes are essentially ``critical'' in nature \cite{Scholzbook,Bak}. A contrasting view of earthquakes has also been widespread, regarding earthquakes as  ``characteristic'' in nature with characteristic energy and time scales \cite{Scholzbook}. 

 In studying statistical  properties of earthquakes, simplified models have been used, {\it e.g.\/}, the so-called spring-block or the Burridge-Knopoff(BK) model \cite{BK,CL89,CL94,MoriKawa1DSR,MoriKawa2DSR,MoriKawaLR,Ohmura}. The OFC model, which was first introduced by Olami, Feder and Christensen (OFC) as a further simplification of the BK model, is one of such statistical models of earthquakes \cite{OFC}. It is a two-dimensional coupled map lattice where the rupture is assumed to propagate from lattice site to its nearest-neighbor sites in a non-conservative manner, often causing multi-site seismic events or ``avalanches''. 

 In the past, many numerical studies have been performed on the OFC model, revealing that the model exhibits certain critical properties such as the GR law \cite{OFC,Grassberger,Lise,Prado,Boulter,Drossel} or the Omori law \cite{Hergarten}. More recent studies also unraveled the characteristic features of the OFC model \cite{Ramos,Kotani,Kawamura}. By investigating the time series of events, Ramos {\it et al\/} found the nearly periodic recurrence of large events \cite{Ramos}. Kotani {\it et al\/} studied the spatiotemporal correlations of the model and identified in the OFC model a phenomenon resembling the ``asperity'' \cite{Kotani}. Physical mechanism underlying such asperity-like phenomena, {\it i.e.\/}, a self-organization process of the stress concentration, was revealed in Ref.\cite{Kawamura}. Thus, the OFC model, though an extremely simplified model, exhibits quite a rich phenomenology containing both critical and characteristic features observed in real seismicity. 

 In the OFC model,  ``stress'' variable $f_i$ ($f_i\geq 0$) is assigned to each site on a square lattice with $L\times L$ sites. Initially,  a random value in the interval [0,1] is assigned to each $f_i$, while $f_i$ is increased with a constant rate uniformly over the lattice until, at a certain site $i$, the $f_i$ value reaches a threshold, $f_c=1$. Then, the site $i$ ``topples'', and a fraction of stress $\alpha f_i$ ($0<\alpha<0.25$) is transmitted to its four nearest neighbors, while $f_i$ itself is reset to zero.  If the stress of some of the neighboring sites $j$ exceeds the threshold, {\it i.e.\/}, $f_j\geq f_c=1$, the site $j$ also topples, distributing a fraction of stress $\alpha f_j$ to its four nearest neighbors. Such a sequence of topplings continues until the stress of all sites on the lattice becomes smaller than the threshold $f_c$. A sequence of toppling events, which is assumed to occur instantaneously, corresponds to one seismic event or an avalanche. After an avalanche, the system goes into an interseismic period where uniform loading of $f$ is resumed, until some of the sites reaches the threshold and the next avalanche starts.  The transmission parameter $\alpha$ measures the extent of non-conservation of the model. The system is conservative for $\alpha =0.25$, and is non-conservative for $\alpha <0.25$. A unit of of time is taken to be the time required to load $f$ from zero to unity.

 It should be noticed that the original OFC model is a spatially homogeneous model, where homogeneity of an earthquake fault is implicitly assumed. Needles to say, real earthquake fault is more or less spatially inhomogeneous, which might play an important role in real seismicity. Then, a natural next step might be to extend the original homogeneous OFC model to the inhomogeneous one where the evolution rule and/or the model parameters are taken to be random from site to site. 

 Possible temporal variation of such a spatial inhomogeneity might also be an important factor. There may be two distinct processes that could change the material parameters characterizing an earthquake fault from one seismic event to the next. The fast dynamical process during an earthquake rupture could change the fault state via, {\it e.g.\/},  wear, frictional heating, melting, {\it etc\/}. In addition, in a long interseismic period until the next earthquake, the fault is subject to many slower processes, {\it e.g.\/}, water migration, plastic deformation, chemical reactions, {\it etc\/},  which would necessarily cause the change in the material parameters at each position on the fault \cite{Scholzbook}. 

 In introducing the spatial inhomogeneity into the OFC model, there might be two extreme ways. In one, one may assume that the randomness is quenched in time, namely, spatial inhomogeneity is fixed over many earthquake recurrences. In the other extreme, spatial inhomogeneity is assumed to vary with time in an uncorrelated way over earthquake recurrences. In the latter extreme, we assign spatial inhomogeneity independently to each successive event, resetting the one of the previous event. In order to get an overview of the role of spatial inhomogeneity in seismicity, full understanding of these two limits of the OFC model would be important and useful.

 There were several previous studies on the inhomogeneous OFC model for both types of inhomogeneities \cite{Ramos,Janosi,Torvund,Ceva,Mousseau,Bach,Jagla}. For the first type of inhomogeneity, {\it i.e.\/}, the quenched or static randomness, Janosi and Kertesz introduced spatial inhomogeneity into the stress threshold and found that the inhomogeneity destroyed the SOC feature of the model \cite{Janosi}. Torvund and Froyland studied the effect of spatial inhomogeneity in the stress threshold, and  observed that the inhomogeneity induced a periodic repetition of system-size avalanches \cite{Torvund}. Ceva introduced defects associated with the transmission parameter $\alpha$, and observed that the SOC feature was robust against small number of defects \cite{Ceva}. Mousseau \cite{Mousseau} and Bach {\it et al\/} \cite{Bach} introduced inhomogeneity into the transmission parameter at each site. These authors observed that the bulk sites fully synchronized  in the form of a system-wide avalanche over a wide parameter range of the model. 

 For the second type of inhomogeneity, {\it i.e.\/}, the dynamical randomness, Ramos considered the randomness associated with the stress threshold, and observed that the nearly periodic recurrence of large events persisted \cite{Ramos}. More recently, Jagla studied the same stress-threshold inhomogeneity, to find that the GR law was washed out by the small amount of randomness \cite{Jagla}.

 In the present paper, we study this second type of inhomogeneity with particular interest in the fate of the asperity-like phenomena identified in the homogeneous OFC model \cite{Kotani,Kawamura}. As to the form of the inhomogeneity, there could be many different implementations. In the present paper, we consider the following three different forms of the spatial inhomogeneity and study their statistical properties by means of numerical simulations.

\medskip\noindent
Model[A]: {\it Random anisotropy in the stress transmission\/}

When the site $i$ topples, a stress $\alpha_1 f_i$ is transmitted to its two nearest-neighbor sites and a stress $\alpha_2 f_i$ is transmitted to the remaining two nearest-neighbor sites. Which neighbor is chosen to be $\alpha_1$ or $\alpha_2$ is decided at random at each site.

\medskip\noindent
Model[B]: {\it Random magnitude in the stress transmission\/}

The transmitted stress is now isotropic among the four nearest neighbors of a toppled site, but its magnitude varies randomly from site to site. We assume that there are two possible values of the magnitude of the stress transmission parameters, $\alpha_+$ and $\alpha_-$. Which $\alpha$ being chosen is determined randomly at each site with the probabilities $p$ and $1-p$, respectively.

\medskip\noindent
Model[C]: {\it Random residual stress value\/}

When the site $i$ topples, the stress  $f_i$ is reset not to zero, but to a finite residual value of $f_r$ ($0<f_r<\delta<1$), where $f_r$ is chosen uniformly between $[0,\delta]$. One may also consider the model where the threshold stress $f_c$ is not unity but distributes around unity (model[C']). Our preliminary study indicates that property of the model[C'] is very much similar to that of model[C]. Hence, in the present paper, we concentrate on the model[C]. 

\medskip
 
 In our simulation, the lattice studied is $L=256$ with open boundary conditions, and the pseudo-sequential updating proposed by Pinho {\it et al\/} being utilized \cite{Pinho}. Initial $1\times 10^8$ avalanches are discarded to reach the steady states, and the subsequent $1 \times 10^9$ avalanches are generated to study statistical properties.

 We begin with the properties of the model [A],  {\it i.e\/}, the inhomogeneous OFC model where the transmission parameter exhibits site-random directional anisotropy, taking a value $\alpha_1$ for two randomly chosen directions and a value $\alpha_2$ for the remaining two directions. This randomness can also be characterized by its mean $\bar \alpha=(\alpha_1 + \alpha_2)/2$ and its standard deviation $\sigma=(\alpha_1 - \alpha_2)/2$.

 We show in Fig.1 the size distribution of avalanches of the model [A] on a log-log plot for several values of the standard deviation  $\sigma$ of the transmission parameter with a fixed mean $\bar \alpha=0.2$. The avalanche size $s$ is defined by the total number of ``topples'' in a given avalanche. As $\sigma$ is varied from 0 (homogeneous model) to 0.015, the SOC feature of the original homogeneous model tends to be weakened, though a near-critical feature still persists to certain extent in the parameter range studied.

 The original homogeneous OFC model exhibits another well-known power-law, the Omori law (the inverse Omori law) associated with the time evolution of the frequency of aftershocks (foreshocks).  A striking contrast to the homogeneous model occurs here. Fig.2 exhibits on a log-log plot the time dependence of the frequency of aftershocks and foreshocks associated with mainshocks of $s\geq s_c=100$ for the case of $\bar \alpha=0.20$ and $\sigma=0.01$. All events of arbitrary size which occur at an arbitrary site on the lattice just after (before) the mainshock of $s\geq s_c=100$ ($t \leq 10^{-4}$) are counted as aftershocks (foreshocks) here. As can be seen from the figure, the introduced inhomogeneity destroys the Omori or the inverse Omori laws almost completely, though a distinct power-law behavior was observed for the homogeneous OFC modelin the same $t$-range of Fig.2 \cite{Hergarten,Kawamura}. Here an increase or decrease of foreshocks/aftershocks itself is washed out by the introduced inhomogeneity.

\begin{figure}[ht]
\begin{center}
\includegraphics[scale=0.7]{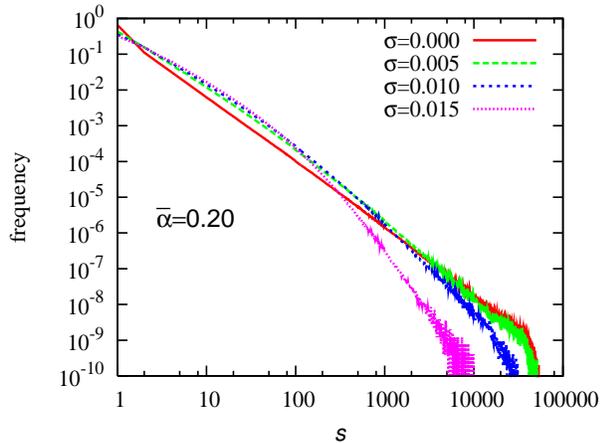}
\end{center}
\caption{
(Color online) Log-log plot of the size distribution of seismic events of the model [A] for various values of the standard deviation $\sigma$ of the transmission parameter with a fixed mean value $\bar \alpha=0.20$. The $\sigma=0$ case corresponds to the homogeneous model.
}
\end{figure}

\begin{figure}[ht]
\begin{center}
\includegraphics[scale=0.7]{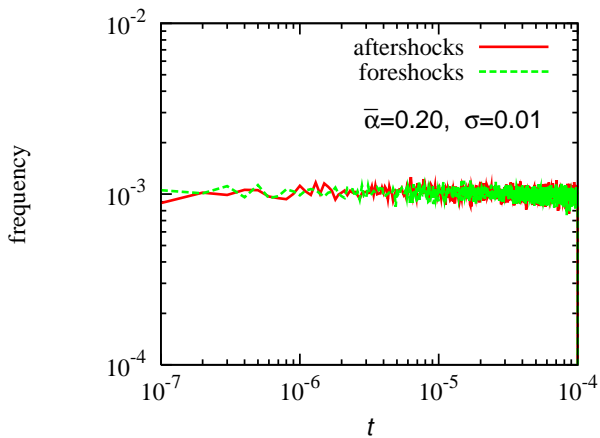}
\end{center}
\caption{
(Color online) The time dependence of the frequency of aftershocks and foreshocks of the model [A] on a log-log plot. The standard deviation of the transmission parameter is $\sigma=0.01$ with its mean $\bar \alpha=0.20$. Mainshocks are the events of their size greater than $s \geq s_c =100$. No range threshold $r_c$ is imposed here in defining foreshocks or aftershocks. The time $t$ is measured with the occurrence of a mainshock as an origin.
}
\end{figure}

 Next, we investigate the local recurrence-time distribution $P(T)$, which turned out to be an efficient probe of the characteristic feature of the model. The local recurrence time $T$ is defined as the  time passed until the next avalanche occurs with its epicenter lying in a vicinity of the preceding avalanche, say, within distance $r\leq r_c$ (in units of lattice spacing) of the epicenter of the preceding event. In the original homogeneous OFC model,  $P(T)$ exhibited a sharp $\delta$-function-like peak at $T=1-4\alpha$ \cite{Kotani}. The computed $P(T)$ of the model[A] is shown in Fig.3 for various  $\sigma$ and fixed $\bar \alpha=0.20$. The size and the range thresholds are taken to be $s\geq s_c=100$ and $r\leq r_{c}=10$. As can be see from the figure, $P(T)$ exhibits a clear peak structure even in the presence of inhomogeneity, indicating that many events of the random OFC model[A] repeat near periodically (the multi-peak structure of $P(T)$ is originated from the size-threshold $s_c$ effect \cite{Kawamura}). As the $\sigma$-value increases, the peak position of $P(T)$ moves to a longer $T$-value, and the width of peak gradually increases: See the inset.

 In Fig.4, we plot the position of the main peak, $T^*$, of $P(T)$ versus $\sigma$ for the cases of $\bar \alpha=0.19$ and $0.20$. As can be seen from the figure, $T^*$ increases with increasing $\sigma$ in proportion to $\sigma$. In fact, the observed $T^*$-value can be well fitted to the form $T^*=(1-4\bar\alpha)(1+5\sigma)$, as shown in Fig.4. For the homogeneous case $\sigma=0$, this expression reduces to the one previously reported for the homogeneous model, $T^*=1-4\alpha$ \cite{Kotani,Kawamura}. The extra factor $1+5\sigma$, which comes from the randomness, is closely related to the stress state of the model, as we shall see below.

\begin{figure}[ht]
\begin{center}
\includegraphics[scale=0.7]{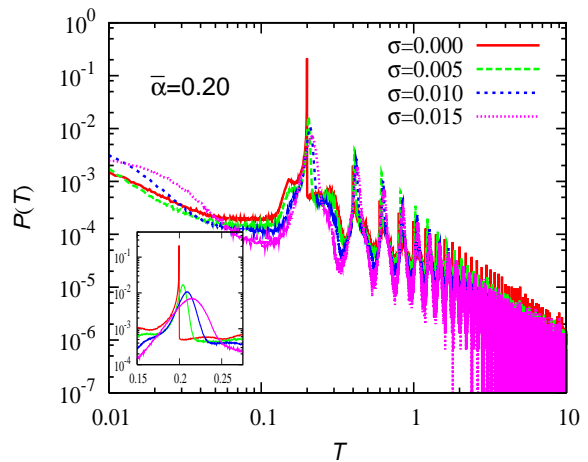}
\end{center}
\caption{
(Color online) Log-log plot of the local recurrence-time distribution of the model [A] for large avalanches of their size $s \geq s_c =100$. The standard deviation of the transmission parameter $\sigma$ is varied from 0 (homogeneous model) to 0.15 with a fixed mean-value of $\bar \alpha=0.2$. The range parameter is $r_c =10$. The inset is a magnified view of the main peak. Note the difference in time scale from that of Fig.2.
}
\end{figure}

\begin{figure}[ht]
\begin{center}
\includegraphics[scale=0.65]{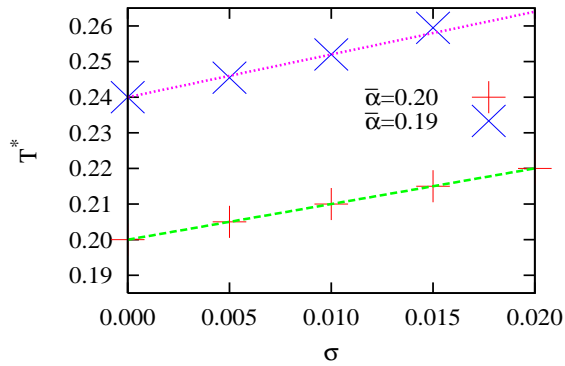}
\end{center}
\caption{
(Color online) The peak position $T^*$ of the local recurrence-time distribution of the model [A] is plotted versus the standard deviation $\sigma$ of the transmission parameter for the mean-values of $\bar \alpha=0.19$ and 0.20. The straight line represents the relation $T^* =(1-4\bar \alpha)(1+5\sigma)$.
}
\end{figure}

 In the original homogeneous OFC model, the sharp structure of $P(T)$ arises due to the near-periodic rupture of the ``asperity-like'' cluster realized in the model \cite{Kotani,Kawamura}, where nearly the same cluster of sites rupture many times near periodically with a period close to $T^*=1-4\alpha$ with almost the same site as its epicenter. We have found that similar asperity-like events occur also in the present inhomogeneous model and are responsible for the sharp peak of $P(T)$. An epicenter site of these asperity-like events lies close to the epicenter site of the preceding asperity event, but often moves from the previous one by several lattice spacings. This might be contrasted to the behavior in the homogeneous model where an epicenter site rarely moves during the asperity sequence \cite{Kotani,Kawamura}. In the homogeneous OFC model, an epicenter site is located at the tip of the rupture zone where only one out of four nearest-neighbor sites is contained in the rupture zone ($n_i=1$ site), while it is never located at the interior site inside the rupture zone ($n_i=4$ site) \cite{Kawamura}. In the present inhomogeneous model, an epicenter site tends to be located at the corner of the rupture zone where two out of four nearest-neighbor sites are contained in the rupture zone  ($n_i=2$ site), while a finite fraction of epicenter sites are located at the interior site inside the rupture zone.

 In Fig.5, we show the stress distribution $D(f)$ at the time of toppling for the sites in the asperity cluster. $D(f)$ exhibits a peak at a stress value $f=f_p$ exceeding the threshold $f_c=1$, whose position depends on the $\sigma$-value.  In the inset of Fig.5,  we plot the observed $f_p$-values as a function of $\sigma$. The data are well fitted to the form $f_p=1+5\sigma$, as shown in the inset. The observation that the asperity site topples at a stress value around $f_p=1+5\sigma$ in the inhomogeneous model, rather than at $f=1$ as in the homogeneous model, might explain the observed deviation of the recurrence time of the asperity events from that of the homogeneous model. Naturally, the random modulation in the transmission parameter $\alpha$ of order $\sigma$ would suppress the formation of the stress state highly concentrated at the threshold $f=1$ by an amount of $\sigma$. Meanwhile, the reason why this factor is given by $1+5\sigma$ quantitatively is not clear to us.

\begin{figure}[ht]
\begin{center}
\includegraphics[scale=0.7]{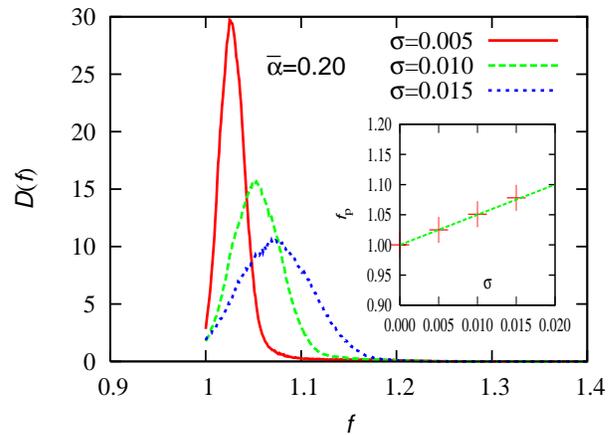}
\end{center}
\caption{
(Color online) The stress distribution $D(f)$ of the model [A] at the time of toppling of each site contained in the rupture zone of the asperity events for several values of the standard deviation $\sigma$ of the transmission parameter. The mean value is fixed to $\bar \alpha=0.2$. The inset represents the peak position of $D(f)$ plotted versus $\sigma$, while the straight line represents the relation $f_p=1+5\sigma$.
}
\end{figure}

 Next, we study the model [B], {\it i.e.\/},  the inhomogeneous OFC model where the transmission parameter, though isotropic in its direction, exhibits a random distribution in its magnitude from site to site, taking a value $\alpha_+$ with the probability $p$ and a value $\alpha_-$ with the probability $1-p$. This randomness can also be characterized by its mean value, $\bar \alpha=p\alpha_+ + (1-p)\alpha_-$, and its standard deviation, $\sigma=\sqrt{p(1-p)}(\alpha_+ - \alpha_-)$. We have studied various cases of $\alpha_+$, $\alpha_-$ and $p$, and have found that the properties of the model [B] is determined by its mean $\bar \alpha$ and standard deviation $\sigma$.

 Indeed, the properties of the model [B] turns out to be very much similar to those of the model [A]. The SOC feature of the size distribution tends to be weakened, while the foreshocks/aftershocks activity is suppressed almost completely. In particular, the local recurrence-time distribution of the model [B] exhibits a clear peak structure borne by the asperity-like events at $T=T^*$ which are very well fittable by the relation $T^*=(1-4\bar \alpha)(1+5\sigma)$, the same formula as we obtained for the model [A]. Hence, the recurrence time of the asperity events of the homogeneous OFC model and the inhomogeneous OFC models [A] and [B] with the random transmission parameter is given by a common simple formula, $T^*=(1-4\bar \alpha)(1+5\sigma)$.  As in the model [A], the stress distribution at the time of toppling for the sites in the asperity cluster also exhibits a peak at a stress value $f=f_p > f_C=1$. Again, the observed $f_p$-values are well fittable by the formula $f_p=1+5\sigma$, explaining the origin of the observed recurrence time of the asperity events, $T^*=(1-4\bar \alpha)(1+5\sigma)$.

%


 Finally, we study the model [C],  {\it i.e\/}, the inhomogeneous OFC model where the residual stress value exhibits a site-random distribution, taking uniformly a value between [0,$\delta$]. The transmission parameter $\alpha$ is assumed to be homogeneous here.  Concerning the avalanche size distribution and  the time evolution of the frequency of foreshocks/aftershocks, a tendency very much similar to the ones of the models [A] and [B] is observed, {\it i.e.\/}, the SOC feature of the size distribution tends to be weakened, while the foreshocks/aftershocks activity is suppressed almost completely.

 Fig.6 exhibits the local recurrence-time distribution of the model [C] for several values of $\delta$ with fixed $\alpha=0.20$.  $P(T)$ exhibits a clear peak structure as in the cases of models [A] and [B], whereas the main peak now lies precisely at $T^*=1-4\alpha$ irrespective of the value of $\delta$, though the peak is broadened with increasing $\delta$. Thus, the randomness in the residual stress does not affect the recurrence time of the asperity events, in contrast to the randomness in the transmission parameter. The peak events are borne by the asperity-like events. An epicenter site here tends to be located either at the corner ($n_i=2$) or at the boundary ($n_i=3$) of the rupture zone.

\begin{figure}[ht]
\begin{center}
\includegraphics[scale=0.7]{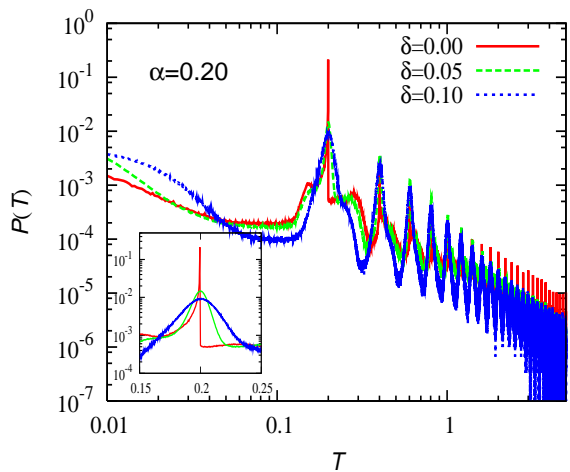}
\end{center}
\caption{
(Color online) Log-log plot of the local recurrence-time distribution of the model [C] for large avalanches of their size $s \geq s_c =100$. The residual-stress threshold $\delta$ is varied from 0 (homogeneous model) to 0.1, while the transmission parameter is taken to be uniform $\alpha=0.2$. The range parameter is $r_c =10$. The inset is a magnified view of the main peak.
}
\end{figure}

 In Fig.7(a), we show the stress distribution $D(f)$ of the model [C] at the time of toppling for the sites in the asperity cluster. $D(f)$ exhibits a peak at a stress value $f=f_p>f_c=1$, whose position depends on the $\delta$-value.  Since the residual stress is nonzero and site-dependent, the distribution of the stress {\it drop\/} $\Delta f$ at each site shown in Fig.7(b), rather than the one of the stress itself at the time of toppling shown in Fig.7(a), might exhibit a peak at $\Delta f=1$. Indeed, this seems to be the case as can be seen from the figure. This observation for the stress drop is fully consistent with our observation in Fig.6 that the recurrence time of the model[C], $T^*=1-4\alpha$, is independent of $\delta$.

\begin{figure}[ht]
\begin{center}
\includegraphics[scale=0.7]{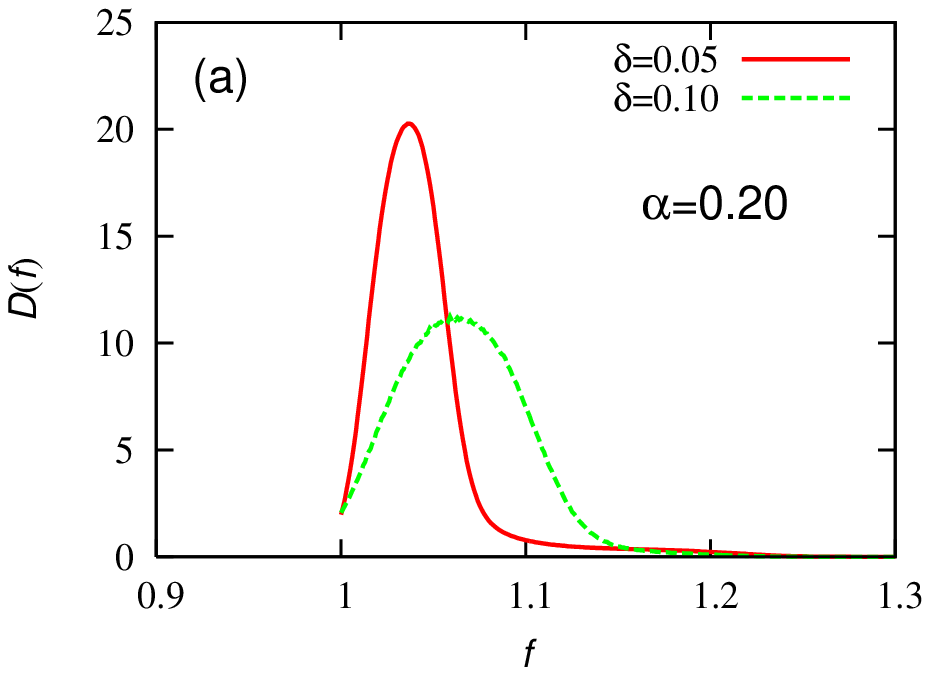}
\includegraphics[scale=0.7]{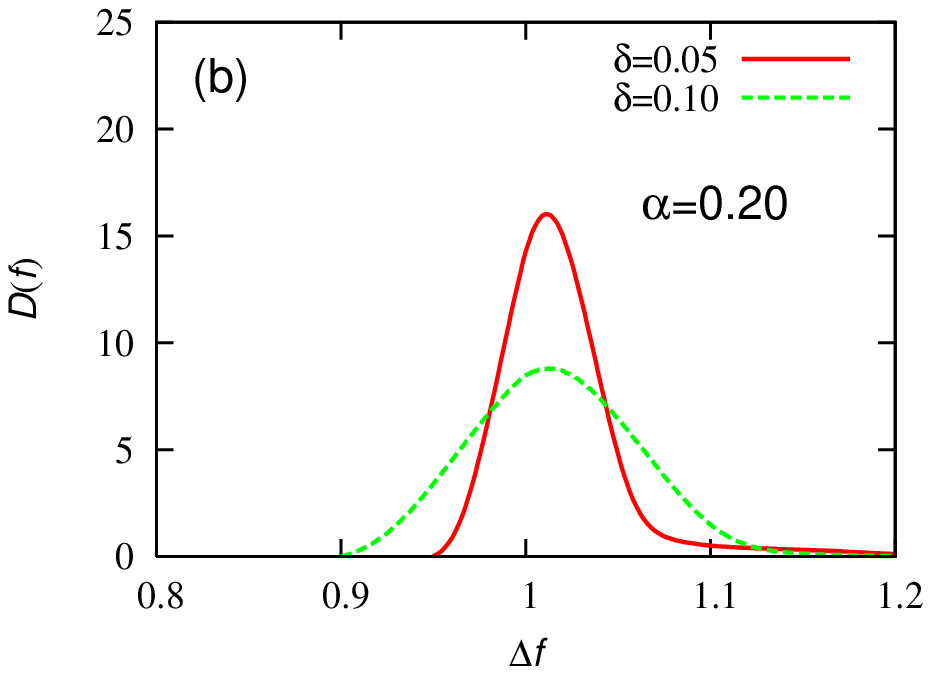}
\end{center}
\caption{
(Color online) The stress distribution $D(f)$ of the model [C] at the time of toppling of each site contained in the rupture zone of the asperity events.  The transmission parameter is $\alpha=0.2$, while the residual-stress threshold is either $\delta=0.05$ or 0.1. In (a), the stress value at the time of the toppling is taken as the abscissa, whereas, in (b), the stress drop, {\it i.e.\/}, the stress value at the time of the toppling minus the residual-stress value at that site, is taken as the abscissa.
}
\end{figure}

 In summary, we studied the statistical properties of the inhomogeneous version of the Olami-Feder-Christensen (OFC) model by numerical simulations. The spatial inhomogeneity was assumed to be dynamical, varying from event to event in an uncorrelated way. In common with all three types dynamical inhomogeneities studied, the critical features of the original homogeneous OFC model are often weakened or suppressed: With increasing the inhomogeneity, the deviation from the GR law becomes more pronounced, accompanied by the suppression of large events.  Our result corroborates the recent observation of Jagla for the model [C'], a model similar to our model [C] \cite{Jagla}. (We note in passing that Jagla further reported that the inclusion of the slow structural-relaxation process in the inhomogeneous OFC model[C'] apparently revived the GR law \cite{Jagla}.). The foreshock/aftershock activity described by the Omori (or inverse Omori) law is entirely gone. By contrast, the characteristic features of the original homogeneous OFC model persist: In all types of dynamical inhomogeneities studied, near-periodic recurrence of large events persist borne by the asperity-like events.  We emphasize that characteristic features are observed for all inhomogeneities in common, suggesting that the persistence of the characteristic features is a generic property of the dynamical inhomogeneity. 
 
 Note also that the properties of the dynamically inhomogeneous models are quite different from those of the static or quenched inhomogeneous models. In the latter case, the introduced inhomogeneity often gives rise to a full synchronization and a periodic repetition of system-size events. Such a system-wide synchronization is never realized in the present dynamically homogeneous models. Presumably, temporal variation of the spatial inhomogeneity may eventually average out the inhomogeneity over many earthquake recurrences, giving rise to the behavior similar to that of the homogeneous model.

This study was supported by Grant-in-Aid for Scientific Research 21540385. We thank ISSP, Tokyo University for providing us with the CPU time.

\end{document}